\documentclass[prb,showpacs,preprintnumbers,amsmath,amssymb,twocolumn,superscriptaddress]{revtex4-1}

\usepackage[dvipdfmx]{graphicx}
\usepackage{dcolumn}
\usepackage{bm}
\usepackage[dvipdfmx]{color}
\usepackage{natbib}

\begin{document}

\title{
Magnetism and Superconductivity in Ferromagnetic Heavy Fermion System UCoGe under In-plane Magnetic Fields
}

\author{Yasuhiro Tada}
\affiliation{Institute for Solid State Physics, The University of Tokyo,
Kashiwa 277-8581, Japan}
\affiliation{Max Planck Institute for the Physic of Complex Systems, 
N{\"o}thnitzer Str. 38, 01187 Dresden, Germany}

\author{Shintaro Takayoshi}
\affiliation{Department of Quantum Matter Physics, University of Geneva,
Geneva 1211, Switzerland}

\author{Satoshi Fujimoto}
\affiliation{Department of Materials Engineering Science, 
Osaka University, Toyonaka, Osaka 560-8531, Japan}

\newcommand{\vecc}[1]{\mbox{\boldmath $#1$}}

\begin{abstract}
We study the ferromagnetic superconductor UCoGe 
at ambient pressure under $ab$-plane
magnetic fields
$\vecc{H}$ which are perpendicular to the ferromagnetic easy axis.
It is shown that, by taking into account 
the Dyaloshinskii-Moriya interaction
arising from the zigzag chain crystal structure of UCoGe,
we can qualitatively explain the experimentally observed in-plane anisotropy
for critical magnetic fields of the paramagnetic transition.
Because of this strong dependence on the magnetic field direction,
upper critical fields of superconductivity, which is mediated by
ferromagnetic spin fluctuations, also become strongly anisotropic.
The experimental observation of ``S-shaped'' $H_{c2}\parallel b$-axis
is qualitatively explained as a result of enhancement of the spin fluctuations
due to decreased
Curie temperature by the $b$-axis magnetic field.
We also show that the S-shaped $H_{c2}$ is accompanied by 
a rotation of the $d$-vector, which would be a key to understand the experiments
not only at ambient pressure but also under pressure. 
\end{abstract}

\pacs{Valid PACS appear here}

\maketitle

\section{introduction}
\label{sec:intro}
Since the discovery of a 
ferromagnetic superconductor UGe$_2$,
a family of ferromagnetic systems,
URhGe and UCoGe, has also been found to exhibit superconductivity
and they have been extensively studied with special focus on 
relationship between ferromagnetism and superconductivity
~\cite{pap:UGe2,pap:Aoki2001,pap:UCoGe2007}.
In these compounds, 5$f$-electrons are responsible both for
the magnetism and the superconductivity,
in sharp contrast to the previously found ferromagnetic superconductors such as 
ErRh$_4$B$_4$ and HoMo$_6$S$_8$ where 
the magnetism and superconductivity have distinct origins~\cite{book:KulicBuzdin}.

Among these uranium compounds, UCoGe has the lowest Curie temperature
$T_{C}\sim 2.7$K and the superconducting transition
temperature $T_{sc}\sim 0.6$K at ambient pressure~\cite{pap:UCoGe2007,
pap:UCoGe2008,pap:UCoGe2009,pap:UCoGe2010,pap:Aoki2009,pap:Aoki2012,pap:Aoki2014}.
The ferromagnetism is suppressed by applying pressure and 
$T_C$ seems to approach zero at a critical pressure $p_c\sim 1.3$GPa.
Below this critical pressure $p<p_c$, 
ferromagnetism and superconductivity coexist in a microspic way
~\cite{pap:UCoGe2009muSR,pap:Ohta2010},
while the superconductivity alone survives up to $p>p_c$.
The experimental 
pressure-temperature phase diagram of UCoGe could be understood from
theoretical model calculations where Ising spin fluctuations mediate
superconductivity~\cite{pap:FayAppel1980,pap:Monthoux1999,pap:Wang2001,pap:Roussev2001,pap:Fujimoto2004}.
Indeed, as revealed by the NMR experiments,
spin fluctuations in UCoGe have strong Ising anisotropy and
the superconductivity is closely correlated with them
especially under magnetic fields~\cite{pap:Ihara2010,pap:Hattori2012,pap:Hattori2014review}.
The experiments show that 
the $a$-axis upper critical field is huge 
$H_{c2}^{\parallel a}>25$T in spite of the low transition temperature
 $T_{sc}\sim0.6$K while $H_{c2}$ for $c$-axis is merely less than 1T,
which leads to cusp-like field angle dependence of $H_{c2}$ in the $ac$-plane.
From a theoretical point of view,
the anomalous behaviors of the observed $ac$-plane upper critical fields of the superconductivity can be
well understood by taking into account the experimental fact that the Ising spin fluctuations
are tuned by a $c$-axis component of the magnetic fields
~\cite{pap:Hattori2012,pap:Tada2013}.
The successful agreement between the experiments and theories
provides strong evidence for a scenario that the pseudo-spin
triplet superconductivity is indeed mediated by the Ising ferromagnetic
spin fluctuations in UCoGe.

On the other hand, different characteristic behaviors have been 
experimentally observed
for $b$-axis magnetic fields in UCoGe
~\cite{pap:Aoki2009,pap:Aoki2012,pap:Aoki2014,pap:Hattori2014}.
In the normal (non-superconducting) states,
the Curie temperature $T_C$ is suppressed by $\vecc{H}\parallel b$-axis and it
seems to become zero around $H^{\ast}\sim 15$T,
although it is unchanged for $\vecc{H}\parallel a$-axis in the same 
experiments. 
The reduction of $T_C$ by $\vecc{H}\parallel b$-axis
is accompanied by an enhancement of the 
spin fluctuations.
Accordingly, at low temperatures,
$H_{c2}$ is {\it enhanced} by the 
$b$-axis magnetic field especially around $H=H^{\ast}$, resulting in 
``S-shaped'' $H_{c2}$.
Interestingly, similar behaviors have also been found in the isomorphic
compound URhGe, where superconductivity vanishes at a critical $\vecc{H}\parallel b$-axis
but it reappears at a high field around which ferromagnetism is suppressed
with a tricritical point~\cite{pap:Levy2005,pap:Tokunaga2015,pap:KHattori2013,
pap:Mineev2014}.

From a theoretical point of view, based on a scenario of the spin
fluctuations-mediated superconductivity,
it is rather natural to expect S-shaped $H_{c2}$ or even reentrant 
superconductivity, once one simply takes into account enhancement of 
the spin fluctuations by the reduction of $T_C$.
However, within this theoretical approach which strongly relies on
the experimental observations of anisotropic behavior
with the application of in-plane magnetic fields,
it is unclear why $T_C$ is unchanged and therefore
$H_{c2}$ is not enhanced for $a$-axis magnetic fields.
In order to understand the dependence of $H_{c2}$ on the direction of 
magnetic field, we should clarify the origin of in-plane magnetic anisotropy.
Furthermore, 
even if one just admits magnetic anisotropy as an experimental fact,
nature of the resulting superconducting state under strong $b$-axis
magnetic field is far from trivial.
For small magnetic fields, the superconductivity will coexist with the
ferromagnetism as in the zero-field case, and it is robust against the Pauli depairing effect
under in-plane magnetic fields due to exchange splitting of the Fermi surface
as pointed out by Mineev~\cite{pap:Mineev2010}.
On the other hand, for larger magnetic fields $H\gtrsim H^{\ast}$ where
the superconductivity survives experimentally,
the exchange splitting is small or even vanishing, and therefore
the Mineev's mechanism protecting the superconductivity from the Pauli depairing
effect does not work.
The limitation of the Mineev's mechanism on the Pauli depairing effect
should also be recognized for understanding $H_{c2}$ under high pressure where
ferromagnetism is suppressed.
Even in paramagnetic states where the Pauli depairing 
effects are expected to be important,
experiments obtain large in-plane $H_{c2}\sim 7-8$T, which is well above the Pauli limiting field
estimated from $T_{sc}\sim 1$K~\cite{pap:Aoki2012review}.
These values of $H_{c2}$ were obtained without fine-tuning of the magnetic field
directions and $H_{c2}$ would be further increased by careful tuning of the field directions,
since it  
sensitively depends on a $c$-axis component of the magnetic fields in UCoGe
~\cite{pap:Aoki2009,pap:Hattori2012}.
Theoretically, it is expected that
the spin fluctuations are large especially around $p=p_c$ leading
to strong coupling superconductivity and the orbital depairing effect
would be less relevant there, while the Pauli depairing effect
is not suppressed and eventually will break the superconductivity.

In this study, we investigate anisotropy for in-plane critical magnetic
fields of paramagnetic transition and superconducting $H_{c2}$ in UCoGe.
Firstly, we make an analysis focusing on zigzag chain crystal
structure which is characteristic in UCoGe.
Within a minimal spin model including effects of the zigzag chain structure,
we show that the Dyaloshinskii-Moriya (DM) interaction arising from the zigzag 
structure leads to the in-plane anisotropy 
for critical magnetic fields of paramagnetic transitions,
which shows a qualitative agreement with the experiments.
We then examine resulting superconducting $H_{c2}$ phenomenologically
considering the suppression of ferromagnetism 
by application of external fields along $b$-axis.
It is shown that superconductivity can survive above $H^{\ast}$
where there is no exchange splitting of the Fermi surface.
We find that this robustness of superconductivity stems from 
the $d$-vector rotation to reduce magnetic energy cost 
as the magnetic field is increased. 
We also touch on the experimental observations based on our calculations.

\section{Anisotropy for critical magnetic field 
of paramagnetic transition and its origin}
\label{sec:FM}
\begin{figure}[htbp]
\begin{center}
\includegraphics[width=0.6\hsize]{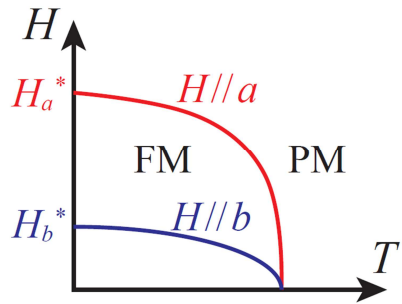}
\caption{Expected magnetic phase diagram of UCoGe in temperature($T$)-magnetic field($H$) plane. 
FM and PM refer to
ferromagnetic phase and paramagnetic phase, respectively.
$H_a^{\ast}$ and $H_b^{\ast}$ are critical magnetic fields at zero temperature.}
\label{fig:mag_phase}
\end{center}
\end{figure}

In this section,
we study the origin of strong dependence of critical magnetic field 
$H^{\ast}$ on the field direction in the $ab$-plane.
As mentioned in the previous section,
the Curie temperature is decreased by magnetic
field along the $b$-axis, while it is unchanged by magnetic field applied 
parallel to the $a$-axis 
within the experimental range 
\cite{pap:Aoki2009,pap:Aoki2012,pap:Aoki2014,pap:Hattori2014}.
If the magnetic fields are further increased, 
$T_C$ will be suppressed for $a$-axis magnetic fields as well.
We schematically show an expected magnetic phase diagram of UCoGe in Fig. \ref{fig:mag_phase}.
$T_C$ decreases rapidly by applying magnetic field along the $b$-axis and eventually
becomes zero at a critical field $H_b^{\ast}$ at zero temperature,
while it is robust against $a$-axis field and the corresponding critical field $H_a^{\ast}$
is much larger than $H_b^{\ast}$.
The purpose of this section is to understand qualitatively 
what causes this anisotropy in the $ab$ plane.
Magnetic anisotropy generally arises from spin-orbit
interactions, and its details depend on strength of the
spin-orbit interactions and crystal structures.
In $f$-electron compounds, 
basic magnetic properties could be well 
understood once the local electronic configuration has been
fixed by e.g. neutron scattering experiments.
For UCoGe, the experimentally observed Ising
magnetic properties may be due to a large weight of $J=5/2$
states in the single electron state at the U-sites.
Although precise determination of a level scheme in 5$f$-electron
systems with low crystal symmetry 
such as UGe$_2$ and UCo(Rh)Ge is very difficult, 
the resistivity in UCoGe shows rather conventional 
heavy fermion behaviors with weak anisotropy in effective mass.
This implies that UCoGe is well described by effective pseudo-spin 1/2 
quasi-particles corresponding to the observed Ising-like magnetism.

\begin{figure}[htbp]
\begin{center}
\includegraphics[width=0.8\hsize]{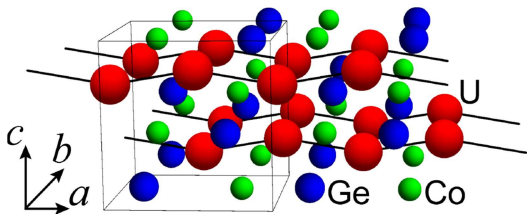}
\caption{The crystal structure of UCoGe.
Zigzag chains are along the $a$-axis.
}
\label{fig:crystal}
\end{center}
\end{figure}

Here, instead of using local electronic structures,
we investigate the magnetic anisotropy in UCoGe by focusing
on its characteristic crystal structure.
As seen in Fig.~\ref{fig:crystal}, 
UCoGe can be viewed as a composition of one-dimensional
zigzag chains along $a$-axis.
The point group of UCoGe is $Pnma$ and the zigzag chains do not have
local inversion symmetry, although they keep global inversion
symmetry.
Such a quasi one-dimensional zigzag structure 
allows an asymmetric spin-orbit (ASO) interaction~\cite{pap:Yanase2014}
\begin{align}
H_{\rm ASO}\sim \sum_{kss'}\sin k_a\sigma^{b}_{ss'}[a_{ks}^{\dagger}a_{ks'}
-b_{ks}^{\dagger}b_{ks'}],
\label{eq:ASO}
\end{align}
where $a_{ks}$ $(b_{ks})$ is an annihilation operator of the quasi-particles
at A (B) sublattice of the zigzag chain.
In order to understand effects of the ASO interaction qualitatively, 
we focus only on spin degrees of freedom 
and introduce a counterpart of the ASO interaction in the spin sector of electrons. 
Then, in terms of the spin degrees of freedom, the ASO interaction 
is mapped to a staggered DM interaction,
\begin{align}
H_{\rm DM}=\sum_{j}(-1)^jD_b[S^c_jS^a_{j+2}
-S^a_jS^c_{j+2}],
\end{align}
where $S_j$ with $j=$odd(even) corresponds to the pseudo-spin 1/2 at the
A(B) sublattice of the zigzag chain.
Note that the direction of the DM vector $\vecc{D}=(0,D_b,0)$
is consistent with a general symmetry argument for $Pnma$ of UCoGe~\cite{pap:Moriya1960};
there is a mirror symmetry with respect to the $(x,1/4,z)$ plane~\cite{pap:bandcalc}, which results in 
$\vecc{D}\parallel b$-axis.
In order to elucidate the effect of this DM interaction 
in the quasi one-dimensional systems,
we investigate a single zigzag chain neglecting inter-chain interactions.
Under this assumption, spin degrees of freedom in UCoGe
is described by the following one-dimensional spin Hamiltonian,
\begin{align}
H_{\rm spin}=-J\sum_jS^c_jS^c_{j+1}-\sum_j[h_aS^a_j+h_bS^b_j]+H_{\rm DM}.
\label{eq:spin}
\end{align}
The first term is the Ising ferromagnetic interaction and the second term corresponds 
to the $ab$-plane magnetic fields.
We have neglected spin-spin interactions of in-plane spin components, 
since the magnetism of UCoGe has strong Ising nature as verified by the experiments
~\cite{pap:Ihara2010,pap:Hattori2012,pap:Hattori2014review}.
In this section, we use a unit where $J=1$.
This spin model should be considered as a variant of the phenomenological Ginzburg-Landau theory 
developed by Mineev, where the free energy is written in terms of magnetic degrees of freedom only
~\cite{pap:Mineev_DM}. 
In the present study, we use the above spin model as an effective phenomenological description to capture 
essential physics behind the complicated experimental results with a special focus on the DM interaction. 
Although
the spin model is oversimplified for discussing quantitative properties of UCoGe,
it is useful for 
qualitative discussions as a minimal model.
Indeed,
as will be discussed in the following, the physical mechanism leading to 
magnetic anisotropy under in-plane magnetic fields revealed within the spin model analysis 
is applicable also for more realistic models of UCoGe.
We also note that, since the Land\'e $g$-factors have not been determined in UCoGe,
the parameters $h_a,h_b$ should be regarded as a renormalized magnetic fields
which include $g$-factors.
Anisotropy of diagonal components of 
the $g$-factors in $a,b$-directions is expected to be small, 
$g_{aa}\simeq g_{bb}$,
since $M$-$H$ curves show weak anisotropy between $M_a$ and $M_b$ 
for small magnetic fields~\cite{pap:UCoGe2008}.
Although there may be off-diagonal components of the $g$-factor 
we simply neglected them.
If $g_{ca}$ or $g_{cb}$ is large, the ferromagnetic phase transitions are smeared out and
become crossover under in-plane magnetic fields.
In the present model without such off-diagonal components,
there is ${\mathbb Z}_2$ symmetry for general $\vecc{h}=(h_a,h_b,0)$
with the operation ${\rm (translation)}\times{\rm (time\mathchar`-reversal)}\times
\exp[i\pi \sum_jS^c_j]$
which transforms spins as
\begin{align}
\left\{
\begin{array}{l}
S^a_j\rightarrow S^a_{j+1},\\
S^b_j\rightarrow S^b_{j+1},\\
S^c_j\rightarrow -S^c_{j+1}.
\end{array}
\right.
\end{align}
This symmetry is spontaneously broken in the ferromagnetic phase.

In order to understand the anisotropy of the critical fields $H^{\ast}$,
we investigate the Hamiltonian \eqref{eq:spin} at $T=0$ by use of
infinite density matrix renormalization group 
(iDMRG)~\cite{pap:DMRG2005,pap:DMRG2011,pap:DMRG2008}. 
In the present one-dimensional model,
we find that
the calculated ground state preserves the translational symmetry and the
system undergoes a quantum phase transition from a uniform ferromagnetic state
with $\langle S^c_j\rangle \neq 0$ to a disordered state with
$\langle S^c_j\rangle = 0$ as $h$ is increased.
We have confirmed absence of a non-uniform magnetic structure by
increasing sizes of assumed sub-lattice structures in the numerical calculations.
This is essentially due to strong Ising anisotropy which favor the colinear ferromagnetic structure,
and coplanar magnetic states might be stabilized if one appropriately 
includes inter-chain coupling and the DM interaction is sufficiently large.
We note that, indeed, such a coplanar state with $a$-axis weak antiferromagnetism
has been predicted within the Ginzburg-Landau theory~\cite{pap:Mineev_DM}.

\begin{figure}
\begin{tabular}{cc}
\begin{minipage}{0.5\hsize}
\begin{center}
\includegraphics[width=0.9\hsize,height=0.7\hsize]{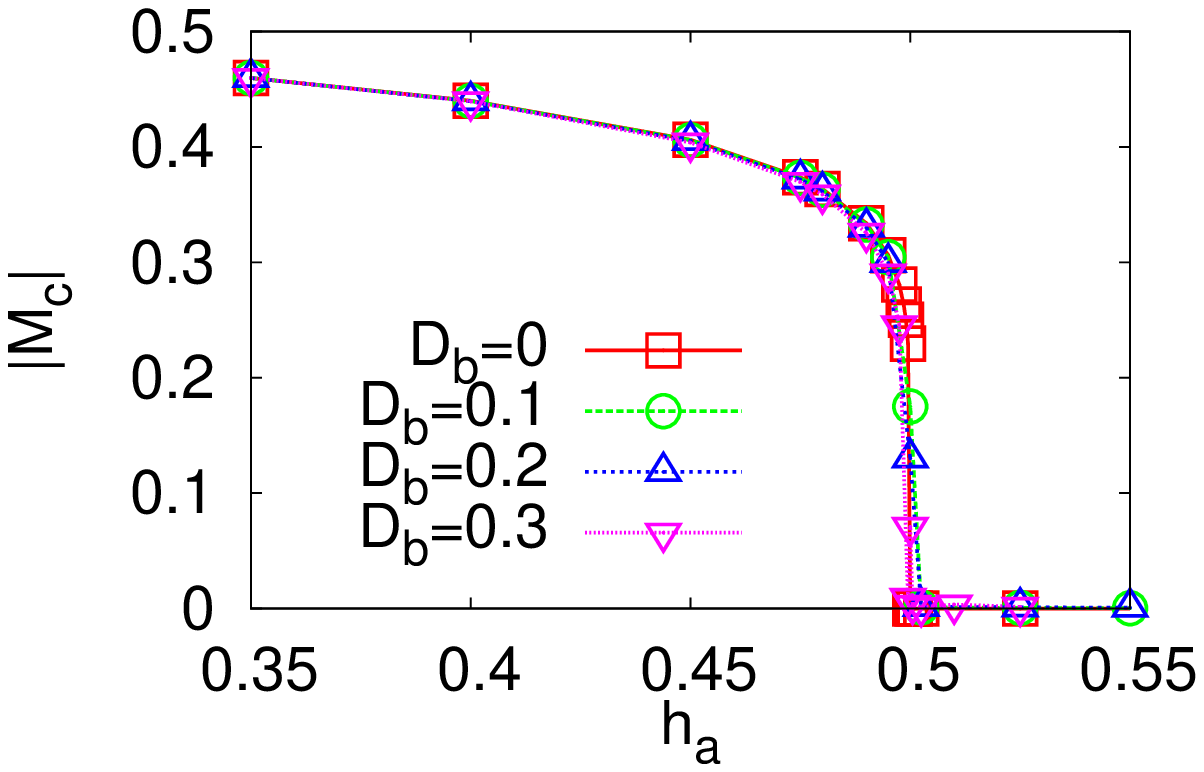}
\end{center}
\end{minipage}
\begin{minipage}{0.5\hsize}
\begin{center}
\includegraphics[width=0.9\hsize,height=0.7\hsize]{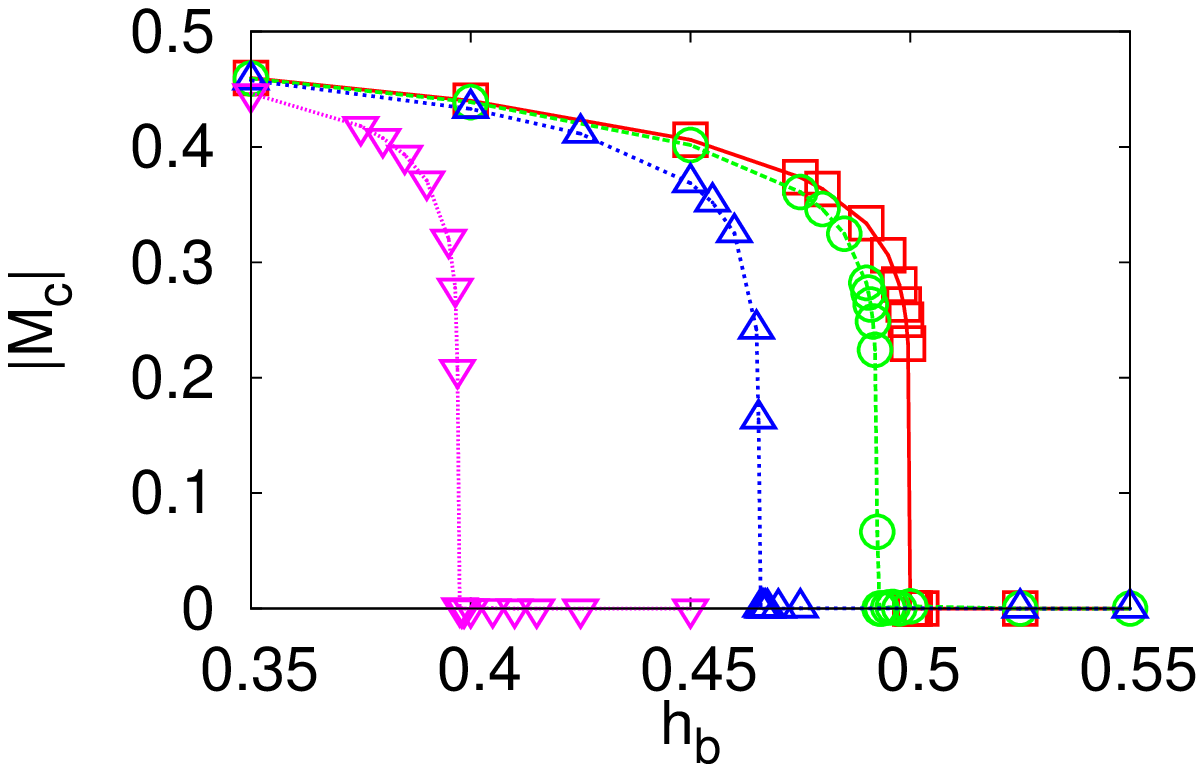}
\end{center}
\end{minipage}
\end{tabular}
\caption{The $c$-axis magnetization $|M_c|$ for different values of the
DM interaction
under the $a$-axis magnetic field (left panel) and $b$-axis magnetic field (right panel).}
\label{fig:mc}
\end{figure}

In Fig.~\ref{fig:mc}, we show magnetization as a function of magnetic fields
for different values of the DM interaction within our model.
For small magnetic fields, the magnetization does not change from that 
without the DM interaction, since the ground state of the Ising Hamiltonian
at $h=0$ is at the same time an eigenstate of
the DM interaction, $H_{\rm DM}|\uparrow\uparrow\cdots\uparrow \rangle=0$.
For large magnetic fields, all the spins are aligned so that
they become parallel to the applied fields.
Interestingly,
the magnetization is not changed by the DM interaction
for $a$-axis magnetic fields even at $h\sim J/2$, 
while it is rapidly suppressed by $b$-axis magnetic fields as the DM interaction is increased.
This anisotropic behavior can be understood as a result of a competition
between the DM interaction and the applied fields;
The DM interaction can be rewritten as~\cite{pap:Oshikawa1997}
\begin{align}
H_{\rm DM}
&=\frac{D_b}{2}\sum_{j:{\rm odd}}[\tilde{S}^+_j\tilde{S}^-_{j+2}
+\tilde{S}^-_j\tilde{S}^+_{j+2}]-(j:{\rm even}),\\
\tilde{S}^{\pm}_j&=e^{\pm i\pi j/4}(S^c_j\pm iS^a_j).
\end{align}
The DM interaction alone describes decoupled
two copies of ``XY-chains'' in the $\tilde{S}$-basis and it
increases quantum fluctuations of spins 
in the $ac$-plane.
Classically, the DM interaction tends to rotate the spins and
it frustrates with the Ising interaction.
However, once strong $a$-axis magnetic fields are applied,
this $ac$-plane quantum fluctuations are pinned and
effects of the DM interactions get suppressed.
Therefore, the calculated magnetization is almost independent of
the DM interaction, and in particular, the transition point almost does not change.
On the other hand,
for $b$-axis magnetic fields $h_b$,
the DM interaction is not suppressed and the ferromagnetic state
is destabilized by the quantum fluctuations, resulting in
smaller critical fields $h_b^{\ast}$.

\begin{figure}
\begin{center}
\includegraphics[width=0.7\hsize,height=0.5\hsize]{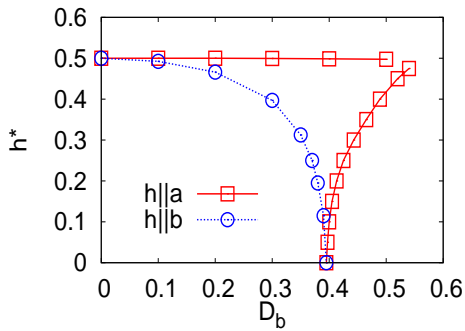}
\caption{The critical magnetic fields
for $a$-axis (red squares)
and $b$-axis (blue circles).
$D_b$ is strength of the DM interaction in unit of $J$.}
\label{fig:Dh}
\end{center}
\end{figure}

We summarize stability of the Ising ferromagnetism against
the DM interaction in Fig. \ref{fig:Dh}.
For $h\parallel a$-axis, the critical field $h_a^{\ast}$ is almost unchanged 
from $h_a^{\ast}\simeq 0.5J$ as $D_b$ is introduced as explained above.
It is noted that the system is dominated by the DM interaction
for large values of $D_b$.
Around $D_b\simeq 0.39J$, there is a first order phase transition between
the Ising ferromagnetic phase and a paramagnetic phase at $h=0$.
The latter state is adiabatically connected to the paramagnetic state
with large $h_a$ and small $D_b$.
On the other hand, for $h\parallel b$-axis,
the critical field $h_b^{\ast}$ is suppressed by $D_b$.
If the DM interaction is sufficiently strong,
the ferromagnetism is more fragile against $h_b$
than $h_a$ at zero temperature. 
This suggests that, at finite temperature, the Curie temperature is
quickly suppressed by $h_b$
compared with $h_a$.
Therefore,
the Hamiltonian \eqref{eq:spin} qualitatively explains
the expected phase diagram Fig. \ref{fig:mag_phase} of UCoGe.
Although these results are based on the simple spin model \eqref{eq:spin},
we believe that the mechanism due to the DM interaction basically applies
to more realistic models.
In general, it is possible that strong magnetic anisotropy remains intact even when one includes 
itinerant nature of electrons into a spin model, although it may be weakened to some extent.
Indeed, UCoGe is an itinerant ferromagnet with strong Ising anisotropy as verified in experiments
~\cite{pap:Ihara2010,pap:Hattori2012,pap:Hattori2014review}.
In order to understand the quantitative features of the magnetic anisotropy
in UCoGe,
one needs to fully include the on-site electron level scheme
together with the ASO interaction.
This issue is left for a future study.

\section{superconducting upper critical field}
\label{sec:SC}
In this section, we consider how the reduced Curie
temperature affects superconducting transition temperature, 
based on the scenario that the 
superconductivity is mediated by the Ising spin
fluctuations in UCoGe.
Similar problems were theoretically studied by several authors
~\cite{pap:KHattori2013,pap:Mineev2014}.
Here, we will focus on qualitative properties and use a simple model
to demonstrate effects of the enhanced spin fluctuations on
the superconductivity.
As was discussed in the previous sections,
low-energy properties in UCoGe can be described by
quasi-particles interacting through
the Ising pseudo-spin fluctuations. 
Therefore, we can approximate our kinetic term as
\begin{align}
S_{\rm kin}&=\sum [i\omega-\varepsilon_k'+\tilde{\vecc{h}}\vecc{\sigma}_{ss'}]
(a^{\dagger}_{ks}a_{ks'}+b^{\dagger}_{ks}b_{ks'})\notag\\
&\quad -\sum \varepsilon_k
(a^{\dagger}_{ks}b_{ks'}+({\rm h.c.}))
+S_{\rm ASO},
\end{align}
where $\varepsilon_k (\varepsilon_k')$ corresponds to inter-sublattice (intra-sublattice)
hopping energy.
The action includes spin-dependent terms described by
$\tilde{\vecc{h}}=\vecc{h}+\vecc{h}_{\rm ex}$, where $\vecc{h}=\mu_B\vecc{H}$ is the applied 
Zeeman field and $g$-factor is simply taken to be $g=2$,
and $\vecc{h}_{\rm ex}$ is the exchange splitting energy of the
Fermi surface in the ferromagnetic state.
From the experiments~\cite{pap:Aoki2009,pap:Hattori2014} and the previous sections,
it is reasonable to assume that
the exchange splitting at zero temperature $\vecc{h}_{\rm ex}\parallel c$-axis
depends only on $b$-axis applied fields.
It is
phenomenologically approximated as
\begin{align}
h_{\rm ex}^c(h_b)=
\left\{
\begin{array}{ll}
h_{\rm ex}(0){\rm tanh}\left(1.74 \sqrt{h_b^{\ast}/h_b-1}\right),&(h_b\leq h_b^{\ast}),\\
0,&(h_b>h_b^{\ast}).
\end{array}
\right.
\label{eq:hex}
\end{align}
This functional form describes a mean field behavior,
$M_c\sim (h_b^{\ast}-h_b)^{1/2}$, near the quantum critical point.
Note that we have neglected $a$-axis applied field dependence of $h_{\rm ex}^c$,
since it is weak as discussed in the previous sections.
One can improve the present model by appropriately modifying $h_{\rm ex}$, e.g.
using the critical exponents of the three dimensional Ising ferromagnets or
introducing temperature dependence.
The kinetic term also includes the ASO interaction Eq. \eqref{eq:ASO} between the 
intra-sublattices.
As in the globally noncentrosymmetric superconductors,
the ASO interaction term tends to fix directions of $d$-vectors for spin-triplet
superconductivity~\cite{book:NCS,pap:Fischer2014}.
In UCoGe, however, Cooper pairing between the nearest neighbor uranium
sites along the 
zigzag chain is expected to be stronger than that between the second
nearest neighbor sites.
The former is inter-sublattice pairing, while the latter is
intra-sublattice pairing.
This suggests that
the ASO interaction between the intra-sublattices 
will affect the sub-dominant gap functions only,
while its effects on the dominant inter-sublattice gap functions would be
negligible in UCoGe. 
Therefore,
we neglect the ASO interaction and do not explicitly take the sublattice structure
into account in the following calculations, which allows us to replace
$a_{ks},b_{ks}$ with a single operator $c_{ks}$.
Then, we use a simple isotropic dispersion
$\varepsilon_k=-2t\sum_{j=a,b,c}\cos k_j, \varepsilon_k'=-\mu$ where
$\mu$ is the chemical potential.
The model parameters are taken to be the same as those in the previous study
~\cite{pap:Hattori2012,pap:Tada2013}, and in particular,
the exchange splitting is $h_{\rm ex}(h_b=0)=0.5t$ which is large enough to 
suppress Pauli depairing effects for small $h_b$.
It should be stressed that the ASO is less important for determining directions of $d$-vector of the 
pseudo-spin triplet pairing between the nearest neighbor sites,
but still relevant to understand the magnetic anisotropy.
The latter effect has already been incorporated in Eq~.\eqref{eq:hex} within the present model
for discussing superconductivity, by neglecting $h_a$ dependence of the exchange splitting..

The fermions interact through
the Ising spin fluctuations which is described by
\begin{align}
&S_{\rm int}=-\frac{2g^2}{3}\sum_q\int_0^{1/T} d\tau d\tau' 
\chi^c(q,\tau-\tau')S_q^c(\tau)S_{-q}^c(\tau'),\\
&\chi^c(q,i\Omega_n)=\frac{\chi_0}{\delta +q^2+|\Omega_n|/\gamma_q},
\end{align}
where $\gamma_q=vq$ with $v=4t$ is the conventional Landau damping factor
and $\vecc{S}_q=(1/2)\sum_k c^{\dagger}_{k-q,s}\vecc{\sigma}_{ss'}c_{ks'}$.
We have neglected interactions arising from in-plane spin components, 
since the Ising spin fluctuations are the dominant fluctuations in UCoGe
\cite{pap:Ihara2010,pap:Hattori2012,pap:Hattori2014review}.
Zero-temperature mass of the Ising spin fluctuations is described by 
$\delta(h_b)$, and
a mean field functional form of $\delta$ is used for simplicity~\cite{pap:FayAppel1980},
\begin{align}
\delta(h_b)=\left\{
\begin{array}{ll}
(h_{\rm ex}^c(h_b))^2 & (h_b\leq h_b^{\ast}),\\
(h_{\rm ex}^c(2h_b^{\ast}-h_b))^2 & (h_b>h_b^{\ast}).
\end{array}
\right.
\end{align}
Since $h_b$-dependence of $\delta(h_b)$ has not been clarified experimentally,
we have assumed that it is symmetric about $h_b=h_b^{\ast}$.
Details of calculation results depend on functional forms of $\delta$,
but their overall behaviors are well captured by this simple function.

In order to calculate $H_{c2}$,
we solve the Eliashberg equation within the lowest order in the
Ising interaction.
The linearized Eliashberg equation reads~\cite{pap:Tada2013,pap:Tada2010}
\begin{align}
\Delta_{ss}(k)&=-\frac{T}{2N}\sum_{k'}V(k,k')[G_{ss'}(k+\Pi)G_{ss'}(-k)\notag\\
&\qquad +G_{ss'}(k)G_{ss'}(-k-\Pi)]\Delta_{s's'}(k'),
\end{align}
where $\vecc{\Pi}=-i\nabla_R-2e\vecc{A}(\vecc{R})$ and $\vecc{A}$
is the vector potential giving a uniform magnetic field.
The pairing interaction $V$ and the selfenergy in the Green's function
$G$ are evaluated as
\begin{align}
V(k,k')&=-\frac{g^2}{6}\chi^c(k-k')+\frac{g^2}{6}\chi^c(k+k'),\\
\Sigma(k)&=\frac{T}{6N}\sum_{qs}g^2\chi^c(q)G_{ss}^0(k+q),
\end{align}
where $G^0$ is the non-interacting Green's function and
$N$ is the number of $k$-point mesh in the Brillouin zone.
We have neglected selfenergies which are off-diagonal in spin space, since they are
much smaller than the diagonal components when $h\ll $ (band width).
In numerical calculations, the lowest Landau level is taken into account 
for the orbital depairing effect.
We focus on the superconducting symmetry for which the $d$-vector is expressed as
$\vecc{d}\sim (c_1k_a+ic_2k_b,c_3k_b+ic_4k_a,0)$ with real coefficients $\{c_j\}$ 
near the $\Gamma$-point
in the Brillouin zone, and the gap function is calculated self consistently
by solving the Eliashberg equation.
We use the same set of model parameters as in the previous studies
~\cite{pap:Hattori2012,pap:Tada2013}, which gives a superconducting transition
temperature $T_{sc0}=0.020t$ at $h=0$.

\begin{figure}
\begin{center}
\includegraphics[width=0.7\hsize,height=0.5\hsize]{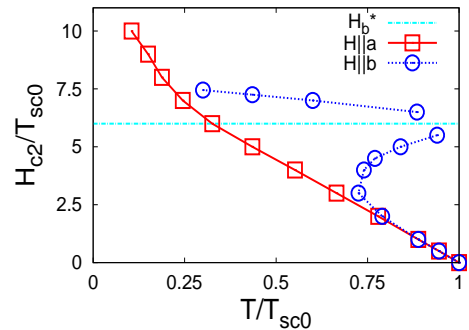}
\caption{Temperature dependence of $H_{c2}$ for $a$-axis (red curve)
and $b$-axis (blue curve).
The dashed line indicates the critical magnetic field $H_{b}^{\ast}$.
$T_{sc0}=0.02t$ is the superconducting transition temperature at $h=0$.}
\label{fig:Hc2}
\end{center}
\end{figure}

In Fig. \ref{fig:Hc2}, we show the upper critical field $H_{c2}^{\parallel b}$
together with the previous results for $H_{c2}^{\parallel a}$
~\cite{pap:Hattori2012,pap:Tada2013}. 
The horizontal light blue line indicates the critical applied field $h_b^{\ast}$ below (above) which
the system is ferromagnetic (paramagnetic).
In our numerical calculations, we cannot accurately compute 
$H_{c2}^{\parallel b}$ near the critical field $h_b\simeq h_b^{\ast}$, since 
strength of the spin fluctuations diverges as $\sim 1/\delta$.
As expected, calculated $H_{c2}^{\parallel b}$ is enhanced around $h_b=h_b^{\ast}$,
which is qualitatively consistent with the experiments on UCoGe~\cite{pap:Aoki2009}.
Although the superconducting transition temperature 
$T_{sc}(h_b=h_b^{\ast})$ seems to exceed $T_{sc}(h_b=0)$ for the present model parameters
in contrast to the experiments,
$T_{sc}(h_b^{\ast})/T_{sc}(0)$ strongly depends on 
$\delta(h_b=0)$ in our model.
If $\delta(0)$ is sufficiently small and the system at zero field is
already very close to the criticality $\delta=0$, enhancement of superconductivity due to
reduction of $\delta(h_b)$ would be moderate~\cite{pap:Monthoux1999,pap:Wang2001,pap:Roussev2001,pap:Fujimoto2004}.
On the other hand, if $\delta(0)$ is rather away from criticality,
enhancement of $T_{sc}$ would be drastic when $\delta(h_b)$ is tuned.
$T_{sc}(h_b^{\ast})/T_{sc}(0)$ also directly depends on the value of $h_b^{\ast}$,
because $b$-axis magnetic fields not only tune the magnetic criticality but also break the superconductivity
at the same time.
However,
it is noted that enhancement of $T_{sc}$ due to the field-induced criticality
is a common qualitative behavior, which is independent of the details.

\begin{figure}
\begin{center}
\includegraphics[width=0.6\hsize,height=0.5\hsize]{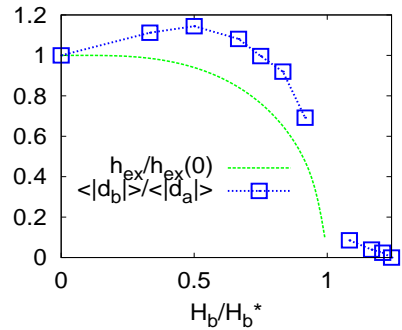}
\caption{In-plane components of the $d$-vector as a function of $H_b$.
The green curve is the exchange field $h_{\rm ex}$ which characterizes
Fermi surface splitting in the ferromagnetic state.}
\label{fig:r}
\end{center}
\end{figure}

It is interesting to see that the superconductivity still survives above the
critical magnetic field, $h_b>h_b^{\ast}$, where the system is paramagnetic.
As was discussed by Mineev~\cite{pap:Mineev2010}, 
a large exchange field $h_{\rm ex}\gg T_{sc}$ is essentially important to 
suppress the Pauli depairing effect for equal-spin pairing states.
Since the equal-spin pairing state is realized and 
$\Delta_{\uparrow\downarrow}=\Delta_{\downarrow\uparrow}=0$
in the present Ising spin fluctuations model,
one might expect that the superconductivity is easily destroyed 
due to the Pauli depairing effect
if the system reaches the paramagnetic state 
with increasing the magnetic field along $b$-axis.
To understand the origin of the robust superconductivity under $b$-axis magnetic fields,
we compare in-plane components of the $d$-vector, $d_a$ and $d_b$,
by calculating 
\begin{align}
\langle |d_a|\rangle&=\sqrt{\frac{1}{N}\sum_k
\frac{|\Delta_{\uparrow\uparrow}(k)-\Delta_{\downarrow\downarrow}(k)|^2}{4}},\\
\langle |d_b|\rangle &=\sqrt{\frac{1}{N}\sum_k
\frac{|\Delta_{\uparrow\uparrow}(k)+\Delta_{\downarrow\downarrow}(k)|^2}{4}}.
\end{align}
Note that absolute values of the $d$-vector cannot be determined within the present calculations
of the linearized Eliashberg equation, but its direction can be self-consistently computed.
We show calculation results in Fig.~\ref{fig:r} together with $h_{\rm ex}(h_b)$ defined in
Eq.~\eqref{eq:hex}.
The calculation results show that $\langle |d_b|\rangle/\langle |d_a|\rangle\simeq 1$ around $h_b\simeq 0$ 
and it goes up when small $h_b$ is introduced.
This is because Cooper pairing only for the Fermi surface of the major spin takes place at $h_b=0$,
and that for the minor spin is induced at finite $h_b>0$.
By further increasing $h_b$, the ratio $\langle |d_b|\rangle/\langle |d_a|\rangle$ 
sharply decreases around $h_b\simeq h_b^{\ast}$
and  it becomes
nearly zero for $h_b>h_b^{\ast}$.
This means that
the $d$-vector at $h_b=0$ is $\vecc{d}\propto (1,i,0)$ and it rotates to
$\vecc{d}\propto(1,0,0)$ for large applied fields which is perpendicular to the applied $b$-axis magnetic field.
The rotation of $d$-vector from the non-unitary state with $\langle |d_a|\rangle
\simeq \langle |d_b|\rangle$ at zero field to the nearly unitary state with 
$\langle |d_a|\rangle\gg \langle |d_b|\rangle$ at large $H_b$ is due to reduction of the exchange splitting of the Fermi surface and 
the Pauli depairing effect. As $H_b$ is increased, the exchange splitting Eq. \eqref{eq:hex} gets smaller, 
which weakens the non-unitarity of $d$-vector. At the same time, in order to reduce Zeeman energy cost, 
the $d$-vector favors a configuration $\vecc{d}\perp \vecc{H}$.
The resulting $d$-vector for $h_b>h_b^{\ast}$ allows a large
Pauli limiting field $H_P^{\parallel b}$ at $T=0$ which is given by
\begin{align}
H_P^{\parallel b}=\Delta_0\sqrt{\frac{\rho(0)}{\chi_N^b-\chi_{sc}^b}},
\end{align}
where $\Delta_0$ and $\rho(0)$ are the gap amplitude and the density of states at the Fermi
energy, respectively.
$\chi_N^b$ is the static susceptibility in normal states
and $\chi_{sc}^b$ is that in superconducting states
given by $\chi_{sc}^b=\chi_N^b[1-\langle\hat{d}_b^2\rangle_{\rm FS}]$ within mean field approximations.
When the $d$-vector is perpendicular to $b$-axis, the susceptibility is $\chi_{sc}^b\simeq \chi_N^b$
and the Pauli limiting field becomes large.
Therefore, the superconductivity can survive up to large $b$-axis magnetic fields 
$h_b\simeq h_b^{\ast}\gg T_{sc}(h_b=0)$ in UCoGe.
However, we note that
the high field superconductivity is numerically stable only for $H_b\gtrsim H_b^{\ast}$ 
and $H_{c2}^{\parallel b}$ is relatively smaller than $H_{c2}^{\parallel a}$ in the present model.
Similar changes of pairing states have been discussed in the previous study for URhGe~\cite{pap:KHattori2013}.

We think that this mechanism for suppressing Pauli depairing effects under in-plane magnetic fields
is important also for the superconductivity under high pressure.
The superconductivity extends over a wide range of pressure and
it survives in the paramagnetic phase~\cite{pap:UCoGe2007,pap:Aoki2009} experimentally.
Although the Mineev's mechanism of suppressing the Pauli depairing effect does not work in the paramagnetic
phase, observed $H_{c2}$ for in-plane magnetic fields are large, 
$H_{c2}\gtrsim$ 7-8T, and they well exceed the
Pauli limiting field $\sim 1$T which is naively expected for the
equal-spin triplet pairing with $T_{sc}\lesssim 1$K~\cite{pap:Aoki2012}.
$H_{c2}$ near the magnetic phase transition point $p_c\simeq 1.3$GPa is especially
non-trivial, since there is a competition between enhanced spin fluctuations and
Pauli depairing effect.
If the $d$-vector rotates to suppress Pauli depairing effect,
the large spin fluctuations lead to strong coupling superconductivity around the critical pressure
and it can be robust against in-plane magnetic fields. 
This mechanism would be relevant for
understanding the observed large $H_{c2}$ in UCoGe under pressure.

\section{summary}
\label{sec:summary}
We have investigated ferromagnetism and superconductivity in the
heavy fermion compound UCoGe under in-plane magnetic fields.
For the magnetic properties,
we focused on roles of the DM interaction
arising from the zigzag chain crystal structure of UCoGe,
and qualitatively explained
the experimentally observed anisotropy for 
the critical field of the paramagnetic transition.
Then we incorporated this magnetic anisotropy into a
simple single-band model 
for the discussion of superconductivity, where 
magnetism is tuned by $b$-axis magnetic fields but is
independent of $a$-axis magnetic fields.
Based on the scenario of the ferromagnetic spin fluctuations
mediated superconductivity, 
we demonstrated that $H_{c2}^{\parallel b}$ shows S-shaped 
behaviors 
in qualitative agreement with the experiments,
while $H_{c2}^{\parallel a}$ is monotonic in temperature.
It was also numerically found that the superconductivity survives even for
large $b$-axis magnetic fields for which
the system is paramagnetic and 
Pauli depairing effect is expected to be relevant.
We showed that the superconductivity survives robustly due to
a rotation of the $d$-vector 
which reduces the Zeeman energy cost
and suppresses the Pauli depairing effect.
The rotation of the $d$-vector would also be important for understanding
large $H_{c2}$ under pressure where the Pauli depairing effect
is not suppressed by 
the exchange splitting of the Fermi surface.

\section*{acknowledgement}
We thank K. Ishida, T. Hattori, D. Aoki, K. Hattori, Y. Yanase, M. Oshikawa, Y. Fuji, and P. Fulde for 
valuable discussions.
Numerical calculations were partially performed by using the supercomputers at
Institute for the Solid State Physics.
This work was supported by JSPS/MEXT Grant-in-Aid for Scientific Research
(Grant No. 26800177, No. 23540406, No. 25220711 and
No. 15H05852 (KAKENHI on Innovative Areas  ``Topological Materials Science"))
and by a Grant-in-Aid for
Program for Advancing Strategic International Networks to
Accelerate the Circulation of Talented Researchers (Grant No.
R2604) ``TopoNet.''


%

\end{document}